Figure 1

| Past | Prediction |
|-------|:----------:|
| 0 0 0 | 1 |
| 0 0 1 | 1 |
| 0 1 0 | 0 |
| 0 1 1 | 1 |
| 1 0 0 | 0 |
| 1 0 1 | 0 |
| 1 1 0 | 0 |
| 1 1 1 | 1 |

# Adaptive Competition, Market Efficiency, Phase Transitions and Spin-Glasses

Robert Savit, Radu Manuca and Rick Riolo
Program for the Study of Complex Systems and Physics Department
University of Michigan
Ann Arbor, MI 48109

## Abstract

In this paper we analyze a simple model of adaptive competition which captures essential features of a variety of adaptive competitive systems in the social and biological sciences. In this model, each of N agents, at each time step of a game, joins one of two groups. The agents in the minority group are awarded a point, while the agents in the majority group get nothing. Each agent has a fixed set of strategies drawn at the beginning of the game from a common pool, and chooses his current best-performing strategy to determine which group to join. We find that for a fixed N, the system exhibits a phase change as a function of the size of the common strategy pool from which the agents initially draw their strategies. For small pool sizes, the system is in an efficient market phase in which all information that can be used by the agents' strategies is traded away, and no agent can accumulate more points than would an agent making random guesses. In addition, in this phase the commons suffer, and relatively few points are awarded to the agents in total. For large initial strategy pool sizes, the system is in an inefficient market phase, in which there is predictive information available to the agents' strategies, and some agents can do better than random at accumulating points. In addition, in this phase, the total number of points awarded to the agents is greater than in a game in which all agents guess randomly, and so the commons do relatively well. At a critical size of the strategy pool marking the cross-over from the efficient market to the inefficient market phases, the commons do best. This critical size of the pool grows monotonically, and in a very simple way with N. The behavior of this system has features reminiscent of a spin-glass in statistical physics, with the small pool size phase being, in a certain sense, more glassy than the large pool phase. We argue that the structure we have elucidated has important implications for a range of phenomena in the social and biological sciences, as well as for the general study of adaptive, competitive systems.

12/3/97 a

**Introduction**

Most systems in the biological and social sciences involve a number of interacting agents, each making behavioral choices in the context of an environment that is formed, in large part, by the collective action of the agents themselves, and with no centralized controller acting to coordinate agent behavior. In the most interesting and difficult to analyze cases, the agents have heterogeneous strategies, expectations and beliefs [Arthur, 1994]. In some cases the agents' strategies may be self-validating, at least for a limited time. For example, in the financial markets a wide-spread belief that a commodity will rise in price may perforce result in a price rise for that commodity [Arthur et al, 1996]. But unless there are fundamental reasons for the price rise, such bubbles eventually tend to burst, so that widely-shared strategies are often self-defeating in the long run. Thus, in many systems, and most clearly in those in which agents compete for scarce resources, successful agents will employ strategies that differentiate them from their competitors, so that the agents will place themselves in groups in which target resources are not over-utilized. Examples of such systems include animals foraging for food, firms searching for profitable technological innovations, packets looking for paths through the internet, or people trying to go to (overly) popular events or places. Furthermore, from the point of view of overall system performance, the best strategy sets are those that result in coordinated resource utilization so that average agent experience is relatively good, and the scarce resource is consumed near its limiting rate. Examples of systems of competing agents in which such coordinated allocation of resources is critical include ecological communities [Cody and Diamond, 1975], routers trying to send packets over the Internet [Kahin and Keller, 1997], and humans trying to decide on which night to go to a popular bar [Arthur, 1994].

These systems are enormously complicated and their detailed dynamics may depend on particular characteristics of the agents and their interactions. Nevertheless, there also are fundamental properties which are shared by all these systems. If we have any hope of ever understanding these kinds of collective adaptive systems, or even of understanding the terms in which we should analyze them, we must first understand the dynamics imposed by their most basic shared properties.

In this paper we analyze a simple model that incorporates the basic adaptive and feedback features of systems of agents competing for a scarce resource [Challet and Zhang, 1997]. In this model, each agent chooses to be in one of two groups at each time step, and those agents in the minority group are awarded a point. Each agent has his own set of strategies, chosen initially from a pool of available strategies, from which he uses the currently best-rated strategy (based on past performance) to select a group to join. Each strategy uses information about which group was in the minority during (a few) previous time steps to predict which will be the minority group during the current time step. We find that as a function of the size of the strategy pool available to the agents, there is a transition which separates an efficient market phase from an inefficient market phase. If the number of available strategies is not sufficiently large, then the system is in a phase, in which the agents are "frustrated" in their attempt to find minority positions, and many features of the aggregate behavior of the system in this phase are analogous to those of the glassy phase of a spin-glass [Mezard, Parisi and Virasoro, 1987; Fischer and Hertz, 1991]. In this phase the market is efficient, and it can be shown that all the relevant information accessible to the agents' strategies has been traded away by their competition [Malkiel, 1985; Fama, 1970; Fama, 1991]. Thus, no agent can ever exceed a 50% rate of success. (Remarkably, however, there is still information in the record of which group was the minority group as a function of time, but that information cannot be used by the agents' strategies.) In addition, the collective behavior of the agents is generally substantially worse than in a random market in which the agents join one of the two groups independently and randomly with equal probability. Above a critical size of the strategy pool, the system is in an inefficient market phase. Here there is predictive information available to the agents' strategies and some agents can achieve better than a 50% success rate. In addition, the collective experience of the agents is much improved and is substantially better than in a random market, indicating an emergent coordination among the agents. As a function of the size of the strategy pool, the best collective performance occurs near the critical size of the strategy pool which marks the cross-over from the efficient market phase to the inefficient market phase. The critical size of the strategy pool scales with N, the number of agents playing the game, in a very simple way. These unexpected emergent properties have profound implications for the study and epistemology of competitive markets in the biological and social sciences.

We first describe the model, then present the results of our analysis and finally discuss the implications of our work for the study of adaptive, competitive social and biological systems.

**The Model**

The simple model of competition we discuss here consists of N agents playing a game [Challet and Zhang, 1997]. The rules of the game are as follows: At each time step of the game, each of the N agents playing the game joins one of two groups, labeled 0 or 1. Each agent that is in the minority group at that time step is awarded a point, while each agent belonging to the majority group gets nothing. An agent chooses which group to join at a given time step based on the prediction of a strategy. The strategy uses information from the historical record of which group was the minority group as a function of time. A strategy of memory m is a table of 2 columns and $2^m$ rows. An example of an m=3 strategy is shown in Fig. 1. The left column contains all the eight possible combinations of three 0's and 1's. To use this strategy, an agent observes which group was the minority group during the last three time steps, and finds that string of 0's and 1's in the left hand column of the table. The corresponding entry in the right hand column contains that strategy's determination of which group (0 or 1) the agent should join during the current time step.

In each of the games discussed here, all strategies used by all the agents have the same value of m. At the beginning of the game each agent is randomly assigned s (>1) of the $2^{2^m}$ possible strategies, with replacement.[1] For his current play the agent chooses his strategy that would have had the best performance over the history of the game up to that time. Ties between strategies are decided by a coin toss. Following each round of decisions, the cumulative performance of each of the agent's strategies is updated by comparing each strategy's latest prediction with the current minority group. Because the agents each have more than one strategy, the game is adaptive in that agents can choose to play different strategies at different moments of the game in response to changes in their environment; that is, in response to new entries in the time series of minority groups as the game proceeds. Because the environment (i.e. the time series of minority groups) is created by the collective action of the agents themselves, this system has very strong feedback, reminiscent of, for example, the financial markets.

**Results**

In what follows we will report and interpret the results of this game for a range of values of N (odd), m and s=2. The qualitative results also hold for other values of s that are not extremely large.[2] To achieve stable results, the game must be run for a long enough time. What is long enough depends on N and m. For our runs, in which N ranged between 11 and 1001, we found that 10,000 time steps were generally sufficient, except for the largest values of N, which required runs of 100,000 time steps. To start the game, we also create a short (of order m) random history of 0's and 1's, so that the strategies can be initially evaluated. The asymptotic statistical results of any run do not materially depend on what this random string is.

To begin to understand the behavior of this system, consider the time series of the number of agents belonging to group 1 ($\equiv L_1$). (This information is not available to the agents but it is available to the researchers.) The mean of this series is generally close to 50% for all values of N, m and s (we shall return to this point below), and so the standard deviation, $\sigma$, of this time series is a measure of how well the commons do: The smaller $\sigma$, the more total points are awarded to all agents combined. That is, if there are typically many fewer than 50% of the agents in the minority, then $\sigma$ will be large and there will be few total points awarded, while if $\sigma$ is small, then most of the time the minority group will consist of only slightly fewer than half of the agents, and more total points will be awarded.

The behavior of $\sigma$ is quite remarkable. In Fig. 2, we plot $\sigma$ for these time series as a function of m for N=101 and s=2. For each value of m, 32 independent runs were performed. The horizontal dashed line in this graph is at the value of $\sigma$ for the random game, i.e. for the game in which the agents randomly choose 0 or 1, independently and with equal probability at each time step.

Note the following features:

1.     For small m, the average value of $\sigma$ is very large (much larger than in the random case). In addition, for m<6 there is a large spread in the $\sigma$ 's for different runs with different (random) initial distributions of strategies to the agents, but with the same m.

2.    There is a minimum in $\sigma$ at m=6 at which $\sigma$ is less than the standard deviation of the random game. We shall refer to the value of m at which the $\sigma$ vs. m curve (for fixed N) has its minimum as $m_c$.[3] Thus, in Fig. 2, $m_c$=6. Also, for m$\geq$$m_c$, the spread in the $\sigma$'s appears to be small relative to the spread for m<$m_c$.

3.    As m increases beyond 6, $\sigma$ slowly increases, and for large m approaches the value for the random game.

The system clearly behaves in a qualitatively different way for small and large m. To further study the dynamics in these two regions, we consider the time series of minority groups, ($\equiv$G) the data publicly available to the agents. We want to study the information content of strings of consecutive elements of this time series of various lengths (including strings of length m) for different values of m and N. To do this, we consider the conditional probability P(1|$u_k$). This is the conditional probability to have a 1 immediately following some specific string, $u_k$, of k elements of G. For example, P(1|0100) is the probability that 1 will be the minority group at some time, given that minority groups for the four previous times were 0,1,0 and 0, in that order. Recall that in a game played with memory m, the strategies use only the information encoded in strings of length m to make their choices. In Fig. 3, we plot P(1|$u_k$) for G, the time series of minority groups generated by a game with m=4, N=101 and s=2. Fig. 3a shows the histogram for k=m=4 and Fig. 3b shows the histogram for k=5. Note that the histogram is flat at 0.5 in Fig. 3a, but, remarkably, is not flat in Fig. 3b. Thus, any agent using strategies with memory (less than or) equal to 4, will find that those strings of minority groups contain no predictive information about which group will be the minority at the next time step. But recall that this time-series was itself generated by players playing strategies with m=4. Therefore, in this sense, the market is efficient and no strategy playing with memory (less than or) equal to 4 can, over the long run, accumulate more points than would be accumulated randomly. (In principle, an agent switching strategies adaptively can do better than random, but we have explicitly verified that this does not happen in this phase. The expicit reason for this is described in the results section and in more detail in [Manuca, Riolo and Savit, 1998].) But note also that the time series of minority groups is not a random (IID)[4] string. There is information in this string, as indicated by the fact that the histogram in Fig. 3b is not flat. However, that information is not available to the agents playing the m=4 game who collectively generated that string in the first place.

We can repeat this analysis for m≥6 (N=101, s=2). For this range of m, the corresponding histogram for k=m is not flat, as we see in Fig. 4 for the m=6 game. In this case, there is significant information available to agents playing the game with memory m and the market is not efficient. Indeed, some individual agents can accumulate more points than they would by simply making random guesses. We have verified these statements by explicitly calculating the points won by individual agents in games with different values of m. For m<6 (N=101, s=2), no agent ever achieves results statistically greater than 50%, while for m≥6 some agents do win more than 50% of the time (statistically significantly). It is noteworthy that even for m=5, the histogram of conditional probabilities with k=m=5 is flat, even though $\sigma$ is less than in the random game.

How does the system behavior change as we change the number of agents? One can repeat the calculation of $\sigma$ for different N. One finds, plotting $\sigma$ vs. m for fixed N, that in all cases one obtains a graph similar to that in Fig. 2, but in which the position of the minimum, $m_c$, is proportional to lnN. In addition, $\sigma$ and the spread in $\sigma$ behave in very simple ways with changes in N which differ depending on whether m is greater than or less than $m_c$. In Fig. 5 we study the behavior of $\sigma$ as a function of N for m=3 and m=16. For the range of values of N used in these figures, m=3 is to the left of the minimum in the curve of $\sigma$ vs. m (i.e. 3<$m_c$ for all these values of N) and m=16 is to the right of the minimum (16>$m_c$). In Fig. 5a we plot $\sigma$ vs. N on a log-log scale. We see that for m=3 $\sigma$ is proportional to N, while for m=16, $\sigma$ is proportional to $N^{1/2}$. This is typical: for fixed m, and m<$m_c$, $\sigma$ is proportional to N, while for fixed m and m>$m_c$ $\sigma$ is proportional to $N^{1/2}$. In Fig. 5b, we plot, again for m=3 and 16, the spread in $\sigma$, i.e., the standard deviation of the $\sigma$'s ($\equiv\Delta\sigma$) as a function of N, on a log-log scale. Here we also see power law behavior: for m=3 $\Delta\sigma$ is proportional to N, while for m=16, $\Delta\sigma$ is proportional to $N^{1/2}$. As before, this behavior is representative of the two behaviors seen for values of m<$m_c$ and values of m>$m_c$, respectively.

The transition between these very different behaviors is at $m_c$~lnN. We have found, using scaling arguments that, to a first approximation, $\sigma^2$/N is a function only of $\frac{2^m}{N}\equiv z$. To see this explicitly, we plot in Fig. 6 $\sigma^2$/N as a function of z on a log-log scale for various N and m (with s=2). We see first that all the data fall on a nearly universal curve. The minimum of this curve is near $2^{m_c}/N=z_c\approx0.5$, and separates the two different phases. The slope for z<$z_c$ approaches -1 for small z, while the slope for z>$z_c$ approaches zero for large z, consistent with the results of Fig. 5.[5]  Because $\sigma^2/N$ depends only on z, it is clear that for fixed z $\sigma$ is proportional to $N^{1/2}$ for any fixed z, both above and below $z_c$.  In addition, it can be shown that, for fixed z $\Delta\sigma$ is approximately independent of N, approaching a z-dependent constant as N→∞.  The N→∞ limit of $\Delta\sigma$ is large for small values of z and decreases monotonically with increasing z.  It is unclear whether or not $\Delta\sigma$ is non analytic at $z_c$.

**Discussion**

In this section, we will first present some qualitative arguments that explain the different behaviors for small and large m, and the scaling results with N.  More detailed explanations along with results of corroborating simulations will appear in a forthcoming publication [Manuca, Riolo, and Savit].  Following that, we will discuss some implications of this study for a wide variety of social and biological systems.

Consider first the small m region (m<$m_c$).  In this region, the time series of the number of agents belonging to group 1 ($\equiv L_1$) has a very unusual structure.  Consider the time series of minority groups, and focus on one particular binary string, $\chi$, of length m.  Now, from $L_1$ form the time series of the number of agents which choose group 1 in response to the first, third, fifth and other odd occurrences of $\chi$.  Now form the time series of the number of agents which choose group 1 in response to all the even occurrences of $\chi$.  It turns out that the standard deviation of the series of responses to the odd occurrences of $\chi$ is of order $N^{1/2}$, while the standard deviation of the time series of responses to the even occurrences of $\chi$ is of order N.  This is true for all possible strings of length m in the time series of minority groups.  Thus, in the whole time series of the number of agents belonging to group 1, there is a bursty structure with large (order N) excursions from the mean separated by smaller excursions of order $N^{1/2}$.  For large N, the contributions of the large, order N excursions to the standard deviation dominate, and so $\sigma \propto N$ in this region.

This behavior can be understood as follows:  Consider the pool of all $2^{2^m}$ strategies.  For s=2, each agent chooses two strategies randomly from the pool.  For very large N, the strategies played by the agents are a good approximation to the population of the pool.  Now, divide the strategies into two groups according to their response to some string, say $\chi$.  The first time $\chi$ appears in the time series of minority groups, the agents will choose to join group 0 or 1, and which strategies they play will be a more or less random sample of the strategies in the pool.  Consequently, we expect that the population of group 1 will deviate from 50% by a number of order $N^{1/2}$. Suppose that the minority group turns out to be group 1. Then, in the pool of strategies, all those whose response to $\chi$ is 1 will get an additional point. The next time the string $\chi$ appears in the series of minority groups, those strategies whose response to $\chi$ is 1 will be preferentially chosen. Now, if N is large, then approximately 75% of the agents will join group 1. (i.e., on average, only those agents both of whose strategies respond to $\chi$ with a 0 will join group 0.) Consequently, the minority group at this time step will be group 0, and all those strategies which respond to $\chi$ with a 0 will get a point. Thus, after the second occurrence of $\chi$, all strategies will have gained one point due to their response to $\chi$, some following the first occurrence, and the others following the second occurrence. There will therefore be no strong preference for the agents' responses to the third occurrence of $\chi$ (but see the discussion in [Manuca, Riolo and Savit, 1998]), and deviations in the membership of group 1 at the third occurrence of $\chi$ will again be of order $N^{1/2}$. This period two process repeats, for all strings and gives rise to the structure described above. A more detailed description of this dynamics will appear in [Manuca, Riolo and Savit, 1998].

The precise distribution of strategies to the agents at the beginning of the game differs from run to run. The existence of the period two oscillations described above does not, for large N, depend on the details of that distribution. However, the coefficient in front of N in the expression for the standard deviation in the responses to even occurrences of a given string does depend on the distribution of initial strategies. Consequently, the size of $\sigma$ will vary from run to run, but, in leading order, will still be proportional to N. Thus, the spread in the $\sigma$'s is also proportional to N in leading order in the low m phase ($m<m_c$), as we saw in Fig. 5.

The existence of different runs with different $\sigma$'s is reminiscent of the glass phase of a spin-glass system. The analogy goes deeper: In the spin-glass case different samples of the glassy material behave differently as a result of frozen in randomly distributed impurities. From one sample to the next, the precise way in which the impurities are distributed affects the thermodynamic behavior of that sample. A spin-glass order parameter, which is non-zero in the glassy phase just expresses the fact that the thermodynamics of different samples is different. In our case, the strategies are distributed to the agents randomly, but the precise way in which that randomness is realized in one game affects the size and distribution of the membership of groups 0 and 1, thus leading to different time-averaged behavior which is most pronounced in the low m phase. Mathematically, $\Delta\sigma$, the spread in the $\sigma$'s, bears some resemblance to a spin-glass order parameter. Furthermore, while the expectation value of the number of agents belonging to group 1 is nearly 50% in the high m phase, it deviates more strongly from 50% in the low m phase, and differs from run to run. This is also analogous to the behavior of a spin-glass, in that the magnetization in the glassy phase is non-zero and differs from sample to sample. Although there is some sense in which the low m phase is glassy, we do not mean to imply that the transition at $m_c$ is a transition from a glassy to a normal phase. Both the high and low m phases partake of glassy behavior and have elements of broken ergodicity due to the initial distribution of strategies, although the glassiness is more pronounced in the low m phase. This analogy will be more fully treated in a forthcoming publication [Manuca, Riolo and Savit].

In contrast to the behavior in the low m phase, for fixed $m \geq m_c$ $\sigma$, the standard deviation of $L_1$, scales with $N^{1/2}$. In this high m phase, for a given N, the strategy space is sufficiently large so that the strategies assigned to the N agents are not a representative sample of the entire strategy space. Consequently, the arguments used in describing the low m phase do not hold, and there are not the same very strong temporal correlations in the response of the system to successive occurrences of a given string. The scaling of $\sigma$ with $N^{1/2}$ just represents the variations in the membership of N agents whose decisions are not very tightly coupled. However, there are still strong dependencies among the decisions of the agents. In fact, it is the remarkable emergent coordination among the agents' decisions that accounts for the fact that $\sigma$ is below the random result in this phase. The precise way in which this comes about will be discussed in detail elsewhere [Manuca, Riolo and Savit, 1998], but the issue can be thought of in the following way: Strategies that would have been most effective in the past at predicting minority groups are more likely to be used by the agents. Those strategies have specific predictions for each string of length m. Consequently, there is an induced dependence among the decisions of the agents which can be expressed as nontrivial conditional probabilities. That is, the probability that at some time agent i chooses group 1, changes depending on the choices of the other agents at that time.[6] It can be shown that, for fixed m, such nontrivial conditional probabilities cause deviations in $\sigma$ from that of the random game. In the high m phase, this term is negative, giving rise to a lower $\sigma$.

Note that the region of greatest coordination (smallest $\sigma$) is when $z = 2^m/N$ is of order one. We can understand this in the following way: Dependencies among the strategies are induced by the agent's selection of the best strategies, as described above. But each chosen strategy dictates a response to $2^m$ different strings of length m. Thus, as m increases, for fixed N, it becomes increasingly difficult for strategies to coordinate all of their entries. (I.e., at any moment there are N strategies in play, and they must minimize fluctuations over $2^m$ choices.) Consequently, for fixed N, maximal coordination will be achieved for some finite m ($=m_c$), whose value is a monotonically increasing function of N. As m increases for fixed N, coordination becomes less effective, and $\sigma$ approaches the result of the random game from below. It can be shown that the correction term is proportional to $1/2^m$ [Manuca, Riolo and Savit].[7]

It is quite remarkable that $\sigma^2/N$ lies on a universal curve as a function of the scaling variable $z=2^m/N$, as shown in Fig. 6. One immediate consequence of this is that the critical value of m, $m_c$ is proportional to lnN as we have found. An intriguing result here is that for maximum coordination N should be roughly the same size as the *dimension* of the strategy space. Thus, it is the dimension of the strategy space that seems to be the relevant measure of the diversity of strategies available to the agents.

**Implications**

This work points up a number of very important and general features that must be confronted in any analysis of adaptive competition among N players. First, it is clear that the size of the strategy space available to the agents and the number of agents playing the game are crucial parameters in determining the qualitative behavior of the system. Moreover, their ratio, z, seems to be the predominant parameter for many of the most important features of the game. In particular, if, for a given number of agents, the strategy space is not large enough, then the game will be efficient in a well-defined sense, but the commons will suffer. There appears to be a critical size of the strategy space that gives rise to the greatest common good, at least as measured by $\sigma$, and everywhere in the inefficient market phase the common good is enhanced over a strictly random game by an implicit emergent coordination. Note that in this game there is no central controlling authority. The improvement of the common good in the inefficient phase is a purely emergent effect. This general structure, an efficient, but poorly performing market when the strategy space is small, a market whose average performance is close to random when the strategy space is large (but which, nonetheless, may produce wealthy agents [Manuca, Riolo and Savit, 1998]), and good performance, both for the commons and for individuals when the dimension of the strategy space is matched to the number of players is likely to be a characteristic that transcends the specific model discussed here. These observations have clear relevance to a number of important issues. For example, there may be significant public policy implications for the design of markets and other regulatory issues, as well as implications for the study of a wide range of competitive systems including ecological systems.

The phase structure of this model as a function of $2^m/N$ is quite intriguing. In particular, it is quite remarkable that there is such a clear transition between the efficient and inefficient phases. Although some of the results of this system may be specific to the model, the observation of the general phase structure and the features it shares with a spin-glass has clear implications for the epistemology of adaptive systems. It suggests, at least as a first step, a set of concepts and tools which are likely to be fruitful in thinking about, and analyzing adaptive competitive systems.

One of our most noteworthy results is that the agents so efficiently and selectively trade away the information accessible to them in the efficient phase ($z<z_c$). The overly competitive dynamics in this phase is remarkably successful at removing information from the reach of the players, and thus removing any possibility that any agent could perform well. What is even more remarkable is the selective way in which this information is removed. There is still significant information contained in the history of minority groups, as we see in Fig. 3b. But it is simply not available to the players whose decisions produced that series in the first place. The precise way in which this information is removed from the system will be discussed elsewhere [Manuca, Riolo and Savit, 1998].

It is clear that the cost of over competition both to the individual and to the collective may be much higher than one would *a priori* have thought. Not only can no agent individually perform well, but the over-competition eliminates the possibility that the common good could be well-served. This may be an important insight for a variety of decision makers. For example, such insights could affect the cost-benefit analysis associated with a company's decision about whether to invest in technological innovation in an already crowded market. There is also an important methodological and conceptual lesson from these observations: The way in which information is removed from access to agents in an efficient or overcrowded market may be considerably more subtle than one might have thought. This suggests a reevaluation of the concept of an efficient market, and the tools and vocabulary used to describe it.

Although the behavior we have elucidated is very intriguing, it is important to remember that there are many effects that may play a major role in specific systems and which could alter the emergent structure, fundamentally. For example, while this model is adaptive, it is not evolutionary. There is no discovery of new strategies by the agents, no mutation, no recombination, no sex. Evolutionary dynamics may drastically alter the phase structure of the system. In fact, we believe that some kinds of evolutionary dynamics may remove the efficient, over competitive phase and drive the system to an effective strategy space corresponding to the region around $m_c$. Nevertheless, any analysis of specific systems which share the competitive dynamics we have discussed here, must take account of the structure we have described, for it is the ground upon which the description of more specific systems, possibly with more complex dynamics, is based.

Finally, and perhaps most importantly, our work raises the question of what really are the fundamental terms in which we ought to think about N-agent adaptive competitive systems. For example, the fact that in the efficient phase ($m < m_c$), $\sigma$ is so strongly dependent on the initial distribution of strategies and, is proportional to N for fixed m, raises the question of how such simple quantities should be interpreted. Clearly, the variance of this time series does not carry the information one might suppose, and at the very least must be supplemented with a specification of the spread in the variance. As another example, the fact that the market is efficient in this phase, but that the time series of minority groups is not random suggests the need for a more sophisticated approach to characterizing publicly available information. The fact is that there is no well developed epistemology for complex adaptive systems, and we are still quite unsure of what the important issues are or what are the most robust ways of characterizing the dynamics of such systems. But the study of simple models, and the elucidation of the variety of behaviors which they manifest can lead us toward a deeper understanding of how to properly frame the questions that we can sensibly ask, and sensibly answer, for complex systems.

**Footnotes**

1.  As we shall see, the dynamics of adaptivity are crucial to our results.  It is, therefore, essential that s>1 so that the agents have more than one strategy with which to play.  For s=1 the game devolves into a game with a trivial periodic structure.  It is also worth noting, parenthetically, that the majority game, in which each agent in the majority group gets a point has trivial periodic structure.

2.  The dependence of the results of the game on s are interesting and will be discussed in detail elsewhere.  However, the qualitative picture we present here obtains for $s << 2^{2^m}$

3.  As we shall see below, the value of m at the minimum of the curve, $m_c$, increases with increasing N.

4.  IID stands for independent and identically distributed, and indicates a sequence whose entries are chosen independently, from some probability distribution that does not change.  In our case, IID would mean that the 0's and 1's were all chosen independently, and that the probability to choose 1 did not change over time.  This is the simplest, most intuitive meaning of "random".

5.  For different values of s there are systematic changes in the shape of this scaling curve. These will be discussed elsewhere [Manuca, Riolo and Savit, 1998].

6.  Another way to say this is that, as a result of the competition among each agent's strategies, the joint probability for any set of agents to make specific choices is not equal to the product of the individual probabilities for each agent's choice.

7.  There is another finite-size effect that contributes to a $\sigma$ that is lower than would be obtained in the random game.  Independent of induced coordination, for a given initial distribution of strategies, the decisions of the N agents following a given string of m 0's and 1's in the time series of minority groups will be constrained.  Therefore, for finite N, the probability distribution of the number of agents choosing group 1 following a specific m-string will not, in general, be symmetric.  It can be shown that skewed distributions lower the standard deviation of a random process.  However, as m increases, the agents will, over time, respond to a larger number of different m-strings, each with a different probability distribution of, say 1's.  Consequently, as m increases, the effect of the skewness in each of the responses to a specific string will be averaged away.

**Figure captions**

Fig. 1.  An example of an m=3 strategy.

Fig. 2.  $\sigma$ as a function of m for N=101 and s=2.  32 independent runs of 10,000 time steps were performed for each value of m.  $\sigma$ for each run is indicated by a dot.  The horizontal dashed line is at the value of $\sigma$ for the random game described in the text. Note the broad spread in the values of $\sigma$ for m<6.

Fig 3a.  A histogram of the conditional probability $P(1|u_k)$ with k=4 for the game played with m=4.  There are 16 bins corresponding to the 16 possible combinations of 4 0's and 1's.  The bin numbers, when written in binary form yield the strings, $u_k$.

Fig. 3b  A histogram of the conditional probability $P(1|u_k)$ with k=5 for the game played with m=4.  There are 32 bins corresponding to the 32 possible combinations of 5 0's and 1's.  The bin numbers, when written in binary form yield the strings, $u_k$.

Fig. 4  A histogram of the conditional probability $P(1|u_k)$ with k=6 for the game played with m=6.  There are 64 bins corresponding to the 64 possible combinations of 6 0's and 1's.  The bin numbers, when written in binary form yield the strings, $u_k$.

fig. 5a.  $\sigma$ as a function of N for fixed m, for two values of m (3 and 16), in the two phases of the system, on a log-log scale.  Note that for m=3, $\sigma \propto N$, while for m=16, $\sigma \propto N^{1/2}$.

Fig. 5b.  The spread in $\sigma$, $\Delta\sigma$, as a function of N for fixed m, for two values of m (3 and 16), in the two phases of the system, on a log-log scale.  Note that for m=3, $\Delta\sigma \propto N$, while for m=16, $\Delta\sigma \propto N^{1/2}$.

Fig 6.  $\sigma^2/N$ as a function of $z = \dfrac{2^m}{N}$ for various values of N, on a log-log scale.

## References

[Arthur, 1994]  W. Brian Arthur, "Complexity in Economic Theory: Inductive Reasoning and Bounded Rationality", Amer. Econ. Assoc. Papers and Proc **84**: 406-411.

[Arthur *et al*, 1996]  W. B. Arthur, J. Holland, B. LeBaron, R. Palmer, and P. Tayler, "Asset Pricing Under Endogenous Expectations in an Artificial Stock Market",  Santa Fe Institute Working Paper 96-12-093.

[Challet and Zhang, 1997]  D. Challet and Y.-C. Zhang, "Emergence of Cooperation and Organization in an Evolutionary Game", preprint, University of Fribourg.

[Cody and Diamond, 1975]  *Ecology and Evolution of Communities*, M. L. Cody and J. M. Diamond (eds.),  (Harvard Univ Press, Cambridge MA).

[Fama, 1970] E. F. Fama,   "Efficient Capital Markets: A Review of Theory and Empirical Work", Journal of Finance; **25**(2), 383-417.

[Fama, 1991] E. F. Fama, "Efficient Capital Markets: II", Journal of Finance, **46**(5), 1575-1617.

[Fischer and Hertz, 1991]  *Spin Glasses*, K. H. Fischer and J. A. Hertz (Cambridge University Press, Cambridge)

[Kahin and Keller, 1997] *Coordination of the Internet*, B. Kahin and J. Keller (eds), (MIT Press, Cambridge).

[Malkiel, 1985],  *A Random Walk Down Wall Street, 4th Edition*, B. G. Malkiel, (W. W. Norton, New York)

[Manuca, Riolo and Savit, 1998] R. Manuca, R. Riolo and R. Savit, in preparation.

[Mezard, Parisi and Virasoro, 1987] *Spin Glass Theory and Beyond*, M. Mezard, G. Parisi and M. A. Virasoro (World Scientific, Singapore).

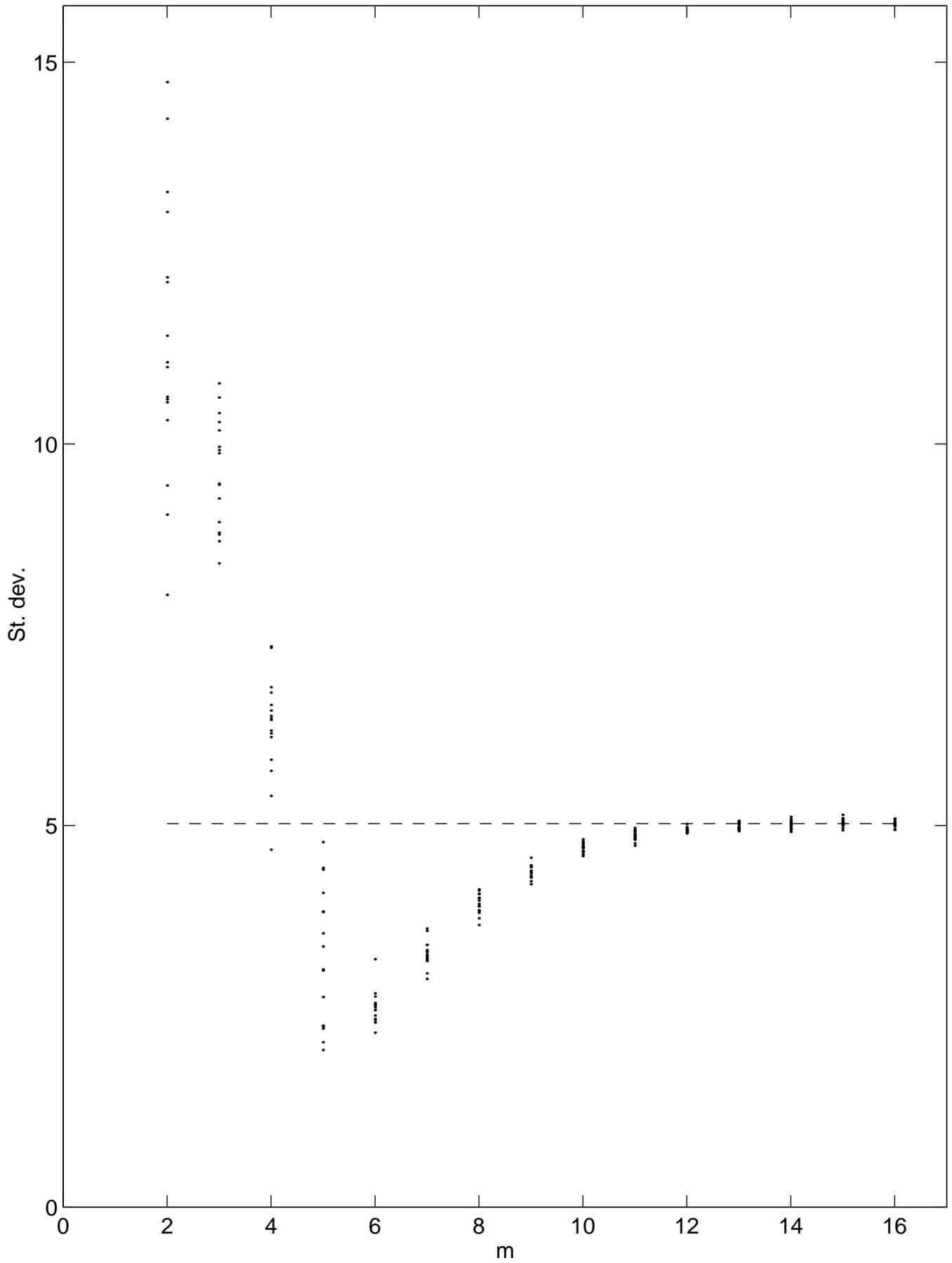

Figure 2: N=101 S=2

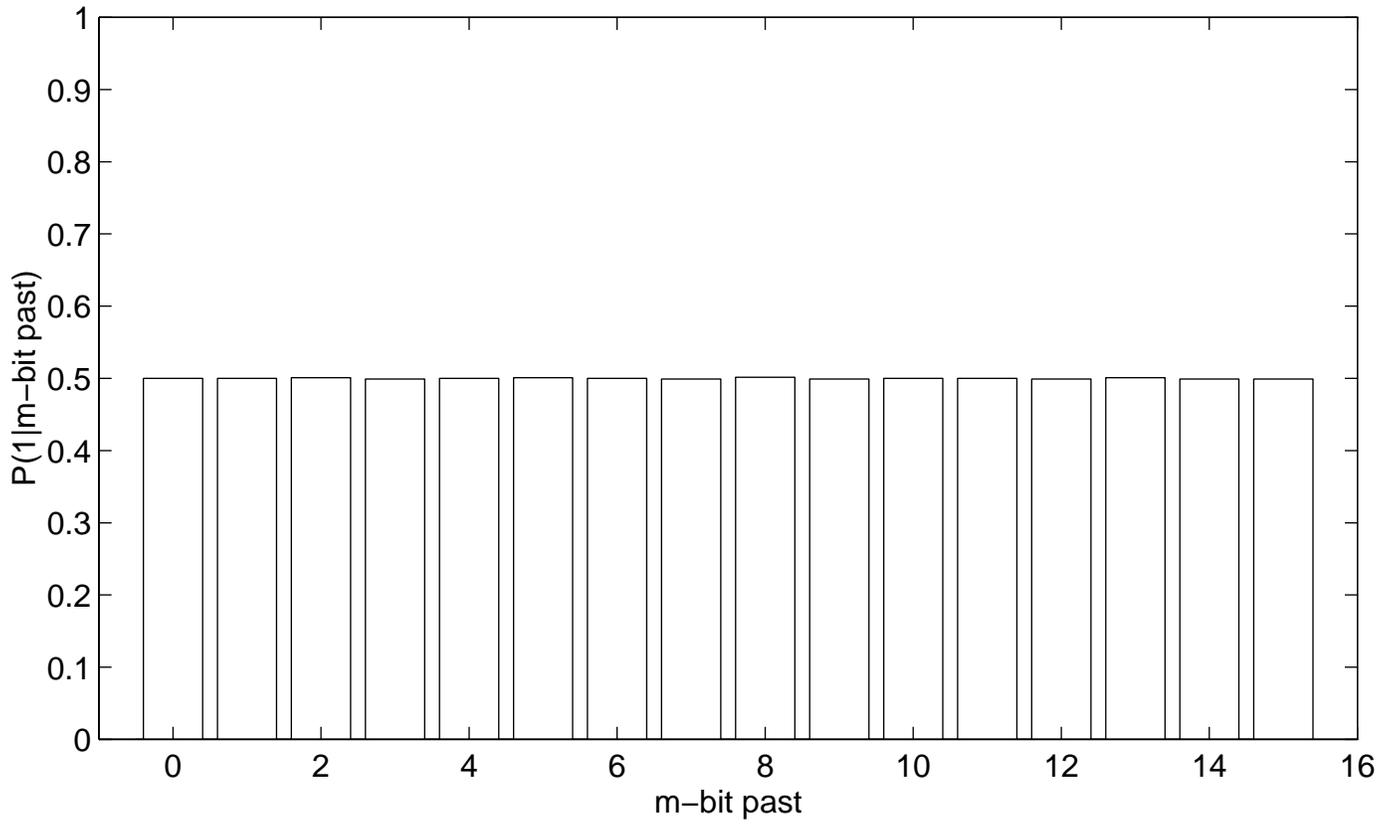

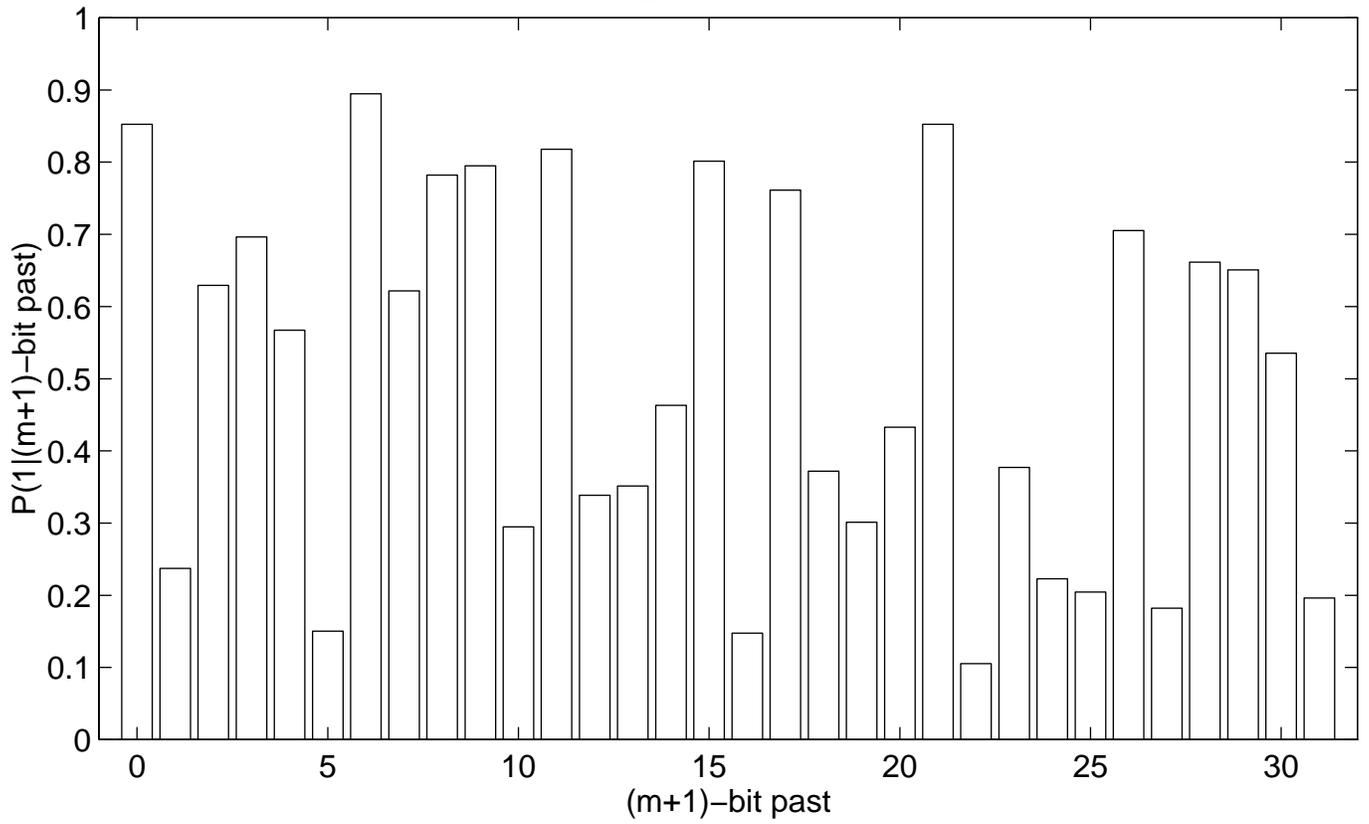

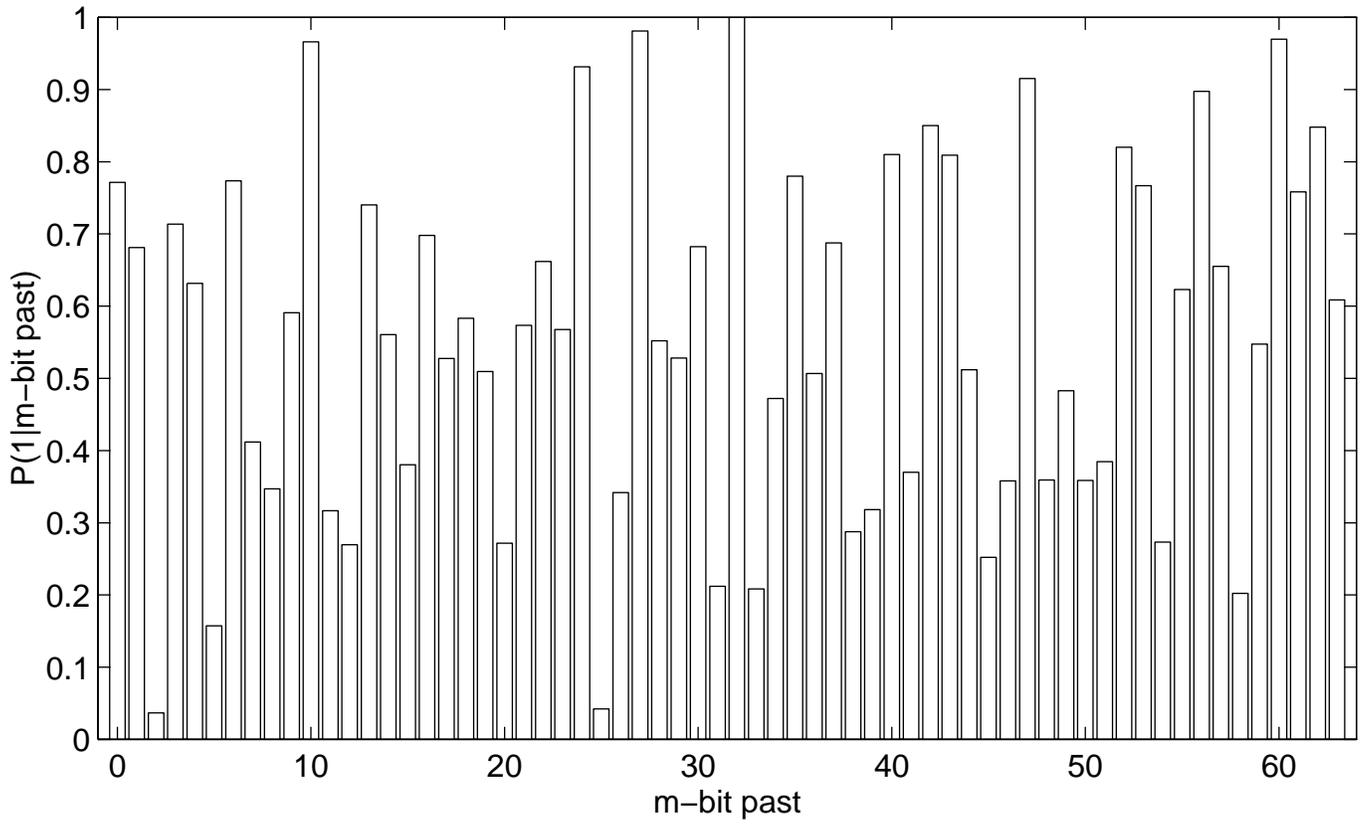

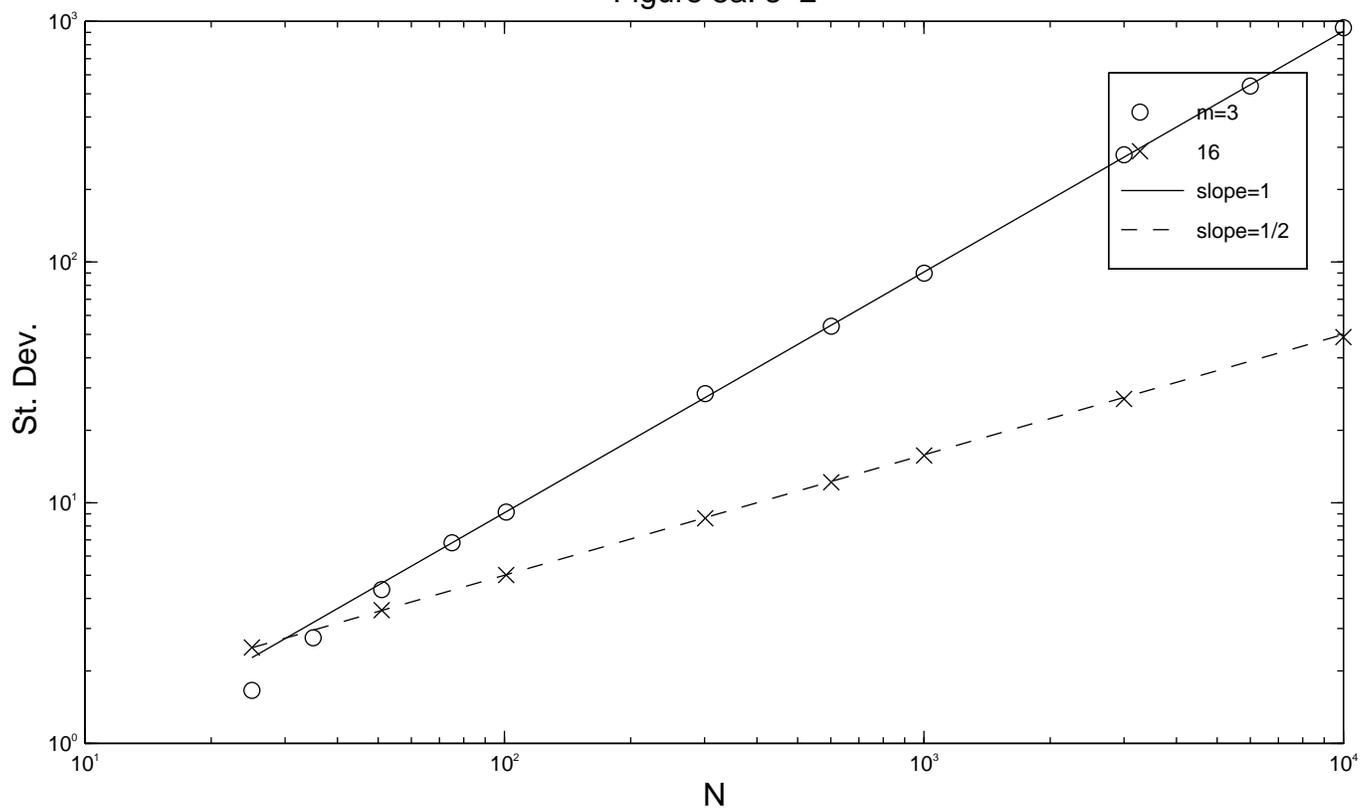

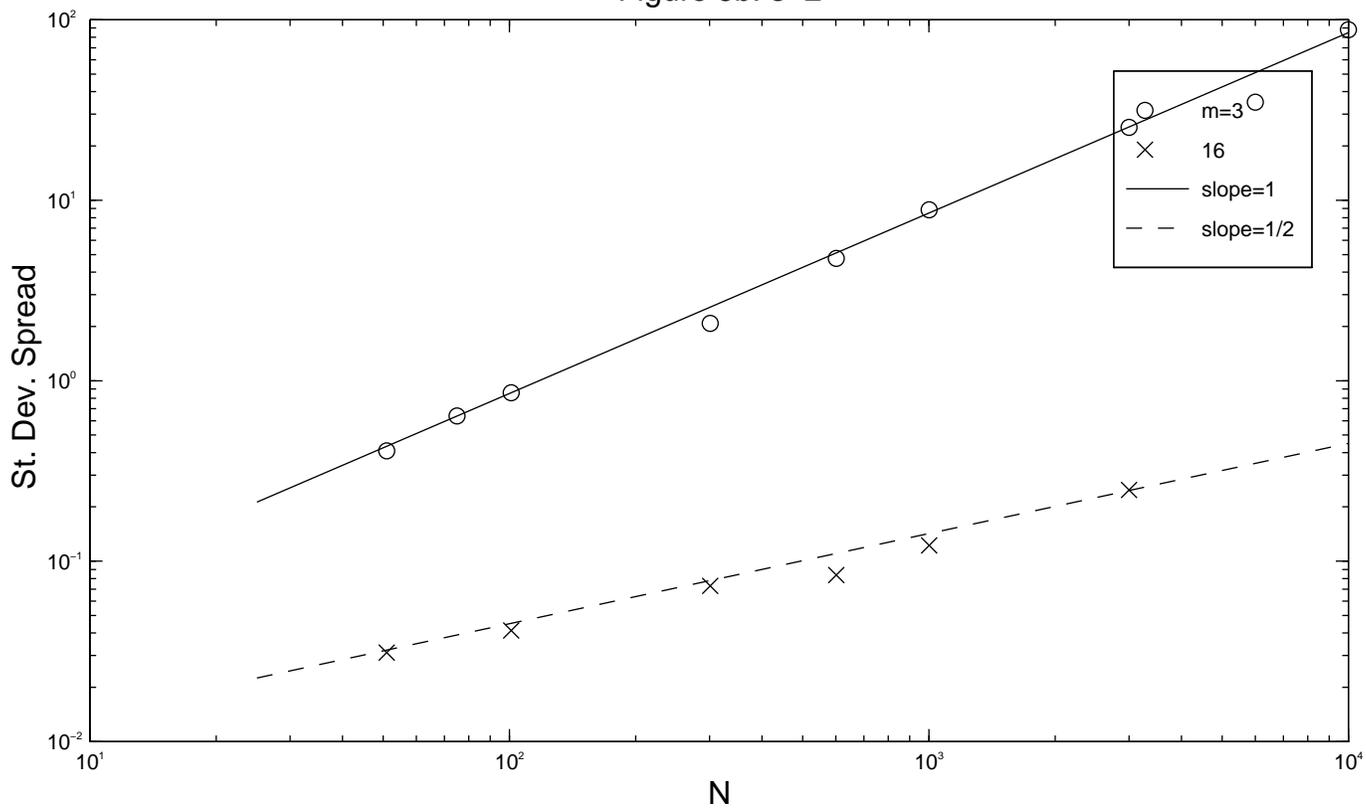

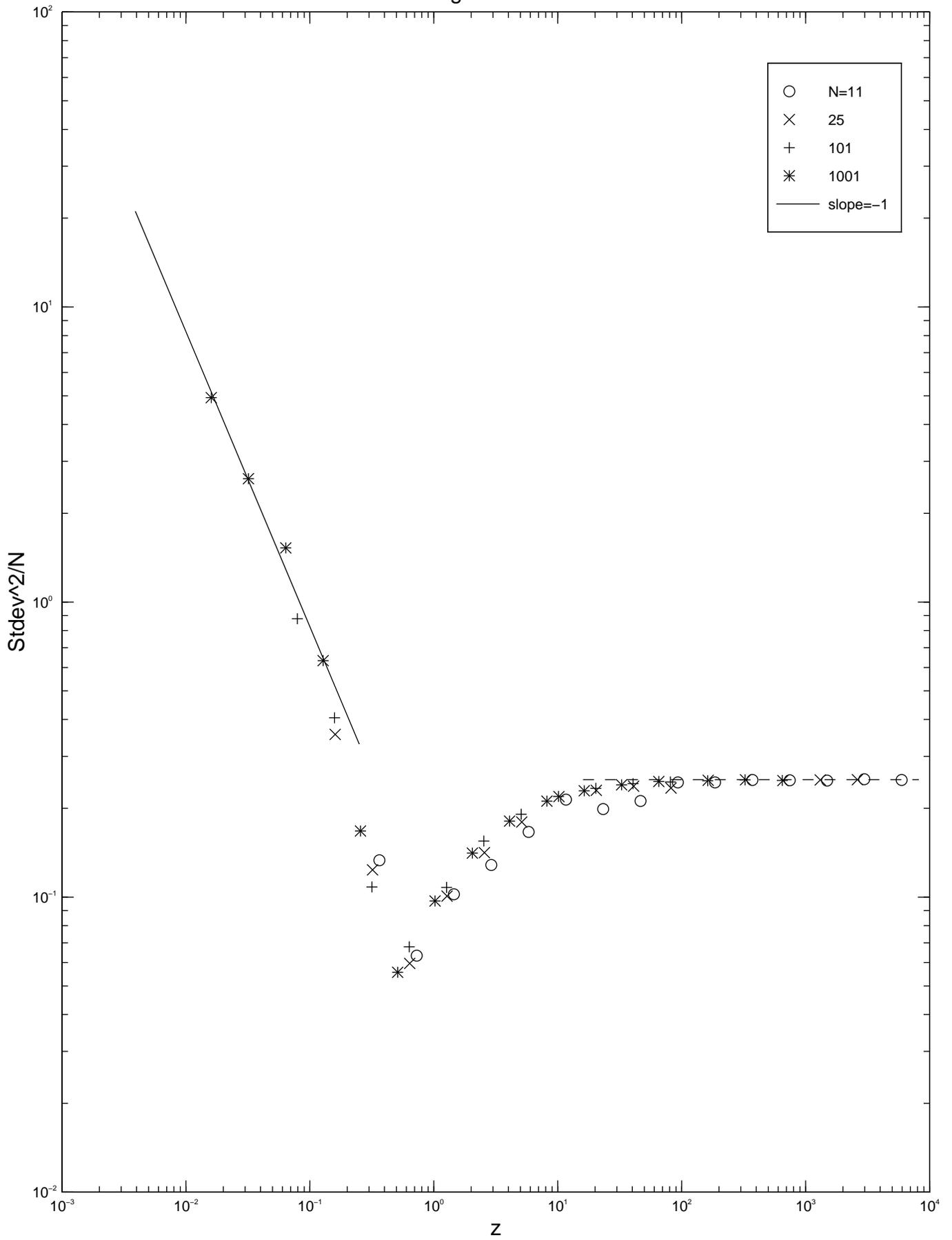